\documentclass[traditabstract]{aa}

\usepackage{graphicx}
\usepackage{txfonts}
\begin{document}

   \title{The VIMOS Ultra-Deep Survey (VUDS): IGM transmission towards galaxies with $2.5<z<5.5$ and the colour selection of high redshift galaxies 
          \thanks{Based on data obtained with the European Southern Observatory Very Large
Telescope, Paranal, Chile, under Large Program 185.A-0791. }
}

\author{R. Thomas\inst{1}
\and O. Le F\`evre\inst{1}
\and V. Le Brun\inst{1}
\and P. Cassata\inst{1,18}
\and B. Garilli\inst{3}
\and B. C. Lemaux \inst{1}
\and D. Maccagni\inst{3}
\and L. Pentericci\inst{4}
\and L.A.M. Tasca\inst{1}
\and G. Zamorani \inst{2}
\and E. Zucca\inst{2}
\and R. Amorin\inst{4}
\and S. Bardelli\inst{2}
\and L. Cassar\`a\inst{3}
\and M. Castellano\inst{4}
\and A. Cimatti\inst{5}
\and O. Cucciati\inst{5,2}
\and A. Durkalec\inst{1}
\and A. Fontana\inst{4}
\and M. Giavalisco\inst{13}
\and A. Grazian\inst{4}
\and N. P. Hathi\inst{1}
\and O. Ilbert\inst{1}
\and S. Paltani\inst{9}
\and B. Ribeiro\inst{1}
\and D. Schaerer\inst{10,8}
\and M. Scodeggio\inst{3}
\and V. Sommariva\inst{5,4}
\and M. Talia\inst{5}
\and L. Tresse\inst{1}
\and E. Vanzella\inst{2}
\and D. Vergani\inst{6,2}
\and P. Capak\inst{12}
\and S. Charlot\inst{7}
\and T. Contini\inst{8}
\and J.G. Cuby\inst{1}
\and S. de la Torre\inst{1}
\and J. Dunlop\inst{16}
\and S. Fotopoulou\inst{9}
\and A. Koekemoer\inst{17}
\and C. L\'opez-Sanjuan\inst{11}
\and Y. Mellier\inst{7}
\and J. Pforr\inst{1}
\and M. Salvato\inst{14}
\and N. Scoville\inst{12}
\and Y. Taniguchi\inst{15}
\and P.W. Wang\inst{1}
}

\institute{Aix Marseille Universit\'e, CNRS, LAM (Laboratoire d'Astrophysique  de Marseille) UMR 7326, 13388, Marseille, France
\and
INAF--Osservatorio Astronomico di Bologna, via Ranzani,1, I-40127, Bologna, Italy
\and
INAF--IASF, via Bassini 15, I-20133,  Milano, Italy
\and
INAF--Osservatorio Astronomico di Roma, via di Frascati 33, I-00040,  Monte Porzio Catone, Italy
\and
University of Bologna, Department of Physics and Astronomy (DIFA), V.le Berti Pichat, 6/2 - 40127, Bologna
\and
INAF--IASF Bologna, via Gobetti 101, I--40129,  Bologna, Italy
\and
Institut d'Astrophysique de Paris, UMR7095 CNRS,
Universit\'e Pierre et Marie Curie, 98 bis Boulevard Arago, 75014
Paris, France
\and
Institut de Recherche en Astrophysique et Plan\'etologie - IRAP, CNRS, UniversitÃ© de Toulouse, UPS-OMP, 14, avenue E. Belin, F31400
Toulouse, France
\and
Department of Astronomy, University of Geneva
ch. d'Ã‰cogia 16, CH-1290 Versoix
\and
Geneva Observatory, University of Geneva, ch. des Maillettes 51, CH-1290 Versoix, Switzerland
\and
Centro de Estudios de F\'isica del Cosmos de Arag\'on, Teruel, Spain
\and
Department of Astronomy, California Institute of Technology, 1200 E. California Blvd., MC 249--17, Pasadena, CA 91125, USA 
\and
Astronomy Department, University of Massachusetts, Amherst, MA 01003, USA 
\and
Max-Planck-Institut f\"ur Extraterrestrische Physik, Postfach 1312, D-85741, Garching bei M\"unchen, Germany
\and
Research Center for Space and Cosmic Evolution, Ehime University, Bunkyo-cho 2-5, Matsuyama 790-8577, Japan 
\and
SUPA, Institute for Astronomy, University of Edinburgh, Royal Observatory, Edinburgh, EH9 3HJ
\and
Space Telescope Science Institute, 3700 San Martin Drive, Baltimore, MD 21218, USA
\and
Instituto de Fisica y Astronom\'ia, Facultad de Ciencias, Universidad de Valpara\'iso, Gran Breta$\rm{\tilde{n}}$a 1111, Playa Ancha, Valpara\'iso Chile
\\ \\
             \email{romain.thomas@lam.fr}
             }

   \date{Received...; accepted...}

 
  \abstract
  {The observed UV rest-frame spectra of distant galaxies are the result of their intrinsic emission combined with absorption along the line of sight produced by the inter-galactic medium (IGM).  
Here we analyse the evolution of the mean IGM transmission $Tr(Ly\alpha)$ and its dispersion along the line of sight for 2127 galaxies with $2.5<z<5.5$ in the VIMOS Ultra Deep Survey (VUDS). We fit model spectra combined with a range of IGM transmission to the galaxy spectra using the spectral fitting algorithm GOSSIP+. We use these fits to derive the mean IGM transmission towards each galaxy
for several redshift slices from $z=2.5$ to $z=5.5$.  We find that the mean IGM transmission defined as $Tr(Ly\alpha)$=$e^{-\tau}$ (with $\tau$ the HI optical depth) is 79\%, 69\%, 59\%, 55\% and 46\% at redshifts 2.75, 3,22, 3.70, 4.23, 4.77, respectively. We compare these results to measurements obtained from quasars lines of sight and find that the IGM transmission towards galaxies is in excellent agreement with quasar values up to redshift z$\sim$4. We find tentative evidence for a higher IGM transmission at z$\geq$4 compared to results from QSOs, but a degeneracy between dust extinction and IGM prevents to draw firm conclusions if the internal dust extinction for star-forming galaxies at z$>$4 takes a mean value significantly in excess of E(B$-$V)$>$0.15. Most importantly, we find a large dispersion of IGM transmission along the lines of sight towards distant galaxies with 68\% of the distribution within 10 to 17\% of the median value in $\delta z$=0.5 bins, similar to what is found on the LOS towards QSOs.
We demonstrate the importance of taking into account this large range of IGM transmission when selecting high redshift galaxies based on their colour properties (e.g. LBG or photometric redshift selection) or otherwise face a significant incompleteness in selecting high redshift galaxy populations. We finally discuss the observed IGM properties and speculate that the large range of observed transmissions might be the result of cosmic variance and clustering along lines of sight, supporting the need for a more complete modelling of the distribution of sources responsible for the extinction. 
}
\authorrunning{Thomas, R., and VUDS team}
\titlerunning{VUDS: the IGM transmission towards galaxies $2.5<z<5.5$}
\maketitle
%


 \section{Introduction}
The light coming from distant sources in the Universe is subject to increasingly 
higher absorption from gas clouds along the line of sight (LOS) as redshift increases. Photons from a distant source with a wavelength corresponding to the Lyman series of Hydrogen at the redshift of the gas cloud are partially or totally absorbed. With many clouds along the LOS, the resulting absorption pattern is known as the Lyman-$\alpha$ forest. As the distance to the source is increasing, the number of gas clouds may become so high that all photons below the Lyman$\alpha$-1216\AA ~line (hereafter, Ly$\alpha$) at the source are totally absorbed. The Ly$\alpha$ forest is thought to be the natural result of hierarchical structure formation within cold dark matter models (e.g. Cen et al. 1994), and is therefore a powerful probe of matter distribution in the Universe.

With a hypothesis on the distribution of gas clouds along the LOS, e.g. their numbers and redshift distribution N(z), it is possible to model the average transmission towards distant sources. The result is quite striking: as shown in models by Madau (1995, hereafter M95), the average transmission is strongly decreasing with increasing redshift and it has a characteristic pattern with several steps corresponding to the Lyman series of Hydrogen, most prominently the Ly$\alpha$, Ly$\beta$, and Ly$\gamma$ transitions. In his models, M95 noted the possible large spread in the average transmission, by computing $1\sigma$ variations from the mean. At $\lambda=1100\AA~$ (rest-frame) the average transmission at $z=3.5$ is predicted to vary from 20\% to 70\%  around a mean of 40\%. The range of LOS absorption was further explored by Bershady et al. (1999) using Monte-Carlo simulations of the IGM absorbers. They find that the range of possible IGM transmission is reduced compared to the model of M95.
Meiksin  (2006, hereafter M06) produced updated IGM models using the $\Lambda$CDM model of Meiksin \& White (2004) and find higher IGM transmission than M95, identified as the result of differences in the estimates of the contributions of resonant absorption. This is confirmed in a recent model by Inoue et al. (2014).

The average IGM transmission $Tr(z)$, has been directly estimated from Ly$\alpha$ forest measurements on the LOS of QSOs. The IGM transmission can be related to $\tau_{eff}$, the HI effective optical depth, as $Tr(Ly\alpha) = e^{-\tau_{eff}}$ to constrain the intensity of the ionizing background (Haardt and Madau 1996, Rauch et al. 1997, Bolton et al. 2005) , and is therefore used to investigate the sources responsible for this ionisation background. Songaila (2004) measure the average IGM transmission and dispersion on the LOS of 50 QSOs and their data show a range of transmission values around the mean. Dall'Aglio et al. (2008) use 40 bright quasars to determine the redshift evolution of the HI effective optical depth in the Ly$\alpha$ forest between 2.2$<$z$<$5\footnote{QSOs studies often consider the redshift $z_\alpha$ defined by $1+z_\alpha=\lambda_0(1+z_s)/\lambda_{Ly\alpha}$ where $z_s$ is the emission redshift of the source (e-g, Becker et al, 2013; Inoue et al, 2014). All the redshifts in our paper are given as $z_s$ and have been transformed from $z_\alpha$ when necessary.}, finding good agreement with measurements based on smaller samples.
Faucher-Gigu\`eres et al. (2008a) report on the mean transmission using 86 QSOs with $2.2 < z < 4.6$ and claim to identify a departure from a power law evolution $\tau_{eff} = 0.0018 (1 + z)^{3.92}$ at around $z\sim3.2$ which they tentatively identify as the signature of the reionisation of HeII. Using 6065 QSOs from the Sloan Digital Sky Survey (SDSS) Becker 
et al. (2013) report on the mean $\tau_{eff}$  over $2.2<z<5.5$, finding no sign of a departure from a smooth evolution with redshift. 
The best measurements indicate an error on the mean transmission of about  $\sim1$\% in $\delta z$=0.1 bins (Becker et al. 2013). 


While the mean value of the IGM transmission is important, the dispersion around the mean is also a key information as the properties of the IGM transmission are extensively used to pre-select galaxies and QSOs in the distant Universe.
In a seminal work, Steidel and collaborators (Steidel et al. 1996) presented the Lyman-break selection
method based on the Lyman-break produced by the Lyman-limit at 912\AA, and its position
in a 3 band filter set producing a colour difference which can  easily be identified from deep photometric observations. 
This galaxy selection has become a standard to find large samples of galaxies, called {\it Lyman Break Galaxies} (LBG),
from a wide range of redshifts. In addition to the 912\AA~ Lyman limit, this technique relies on the average IGM transmission, as the observed flux beween the Lyman limit and Ly$\alpha$ is strongly affected by the IGM. At increasing redshifts, as the IGM extinction gets higher, a strong apparent continuum break builds-up as flux blueward of Ly$\alpha$ at the source is increasingly suppressed. This property is extensively used to identify {\it dropout} galaxies up to the highest possible redshifts (e.g. Bouwens et al. 2014 and references therein). A good knowledge of the average IGM transmission and the dispersion around the mean is then not only crucial to understand the properties of the IGM itself, with the distribution and content of the gas clouds, but also to understand how high redshift galaxies are selected and eventually missed. 

Today the knowledge of IGM transmission is mostly coming from simulations (M95, M06, Inoue et al. 2014), coupled to observational evidence of the average IGM transmission $Tr(Ly\alpha)$, or equivalently the effective optical depth $\tau_{eff}$, coming from the analysis of QSO lines of sight (Becker et al. 2013 and references therein). 
Surprisingly, there are only few reports on the observed dispersion in $Tr(Ly\alpha)$ as function of redshift. Faucher-Gigu\`eres et al. (2008b) use 86 high-resolution, high signal-to-noise quasar spectra to provide reference measurements of the dispersion in $Tr(Ly\alpha)$ over 2.2$<$z$<$4.6. Cutting the IGM along the LOS in 3 Mpc segments they present the observed $Tr(Ly\alpha)$ for these individual measurements, with a standard deviation among the individual 3Mpc segment of $\sigma(Tr(Ly\alpha))_i$=0.1--$0.15$. The individual $Tr(Ly\alpha)$ reported in Dall'Aglio et al. (2008)  indicate a similar dispersion.

There is no observational study for the evolution of IGM transmission from galaxy samples, mainly because of the lack of large spectroscopic samples with high signal-to-noise probing significantly bluer than Ly$\alpha$, and hence the comparison of IGM transmission towards extended galaxies with point-like QSOs has not yet been performed. 
In this paper we use  2127 highest signal to noise galaxies in the VIMOS Ultra Deep Survey (VUDS; Le F\`evre et al. 2014) to compute the mean IGM transmission, and the distribution around the mean, in the range $2.5<z<5.5$. This large sample enables to probe a large number of different lines of sight, and hence to characterize the statistical properties of the IGM transmission towards galaxies. We describe the VUDS galaxy sample in Section \ref{data}. The GOSSIP+ spectral fitting algorithm and the range of IGM templates used in spectral fitting is described in Section \ref{measure_fit}. The results of the spectral fitting on IGM transmission and dispersion are described in Section \ref{results}. The evolution of the Ly$\alpha$ effective optical depth is presented in Section \ref{teff}. We discuss the uncertainties in measurements and models in Section \ref{discuss}. The impact of the IGM transmission dispersion on the selection of high redshift galaxies is analysed in Section \ref{selec}. Our results are summarized in Section \ref{summary}.

All magnitudes are given in the AB system unless specified,
and we use a Cosmology with $\Omega_M=0.3$,
$\Omega_{\Lambda}=0.7$ and $h=0.7$.  
 
\section{Data: the VUDS survey}
\label{data}

The VUDS survey is described in details in Le F\`evre et al. (2014); we provide a summary in this section. The main target selection of the survey is based on photometric redshifts with $z_{phot} + 1\sigma \geq 2.4$, and $i_{AB} \leq 25$, and is also including objects for which the secondary peak in the photometric redshifts probability distribution function is $z_{phot} + 1\sigma \geq 2.4$ even if the primary peak might be very different. This primary target selection is supplemented by colour-colour selection (BzK, LBG) adding those objects which failed satisfying the $z_{phot}$ criterion, but which satisfy the colour-colour selection at the corresponding redshift, adding about 10\% of objects. Observations are performed with the VIMOS multi-slit spectrograph on the ESO-VLT (Le F\`evre et al. 2003), using both the LRBLUE and LRRED grisms with a spectral resolution $R \simeq 230$, and integration times of $\sim$14h for each grism, covering a wavelength range $3650 \leq \lambda \leq 9350$\AA. 

Data were processed within the VIPGI environment (Scodeggio et al. 2005), and redshifts measured with the EZ code (Garilli et al. 2010) based on cross-correlation with reference templates. The redshift measurement process includes visual inspection of all spectra by two independent observers, each making their best redshift measurement running EZ manually if necessary, and these two measurements are confronted in a face-to-face meeting to produce the final redshift measurement and associated reliability flag. All VUDS galaxies are matched to deep photometric samples existing in each of the 3 VUDS fields: COSMOS, ECDFS and VVDS-02h, as described in Le F\`evre et al. (2014). 

To study the absolute amount of IGM transmission in observed spectra, the UV flux has to be calibrated to better than 10\% over the observed wavelength range, particularly for all wavelengths below Ly$\alpha$. The relative spectrophotometry  over the entire VIMOS wavelength range calibrated on reference stars is accurate at better than a few percent level (Le F\`evre et al. 2003) when observations are taken at the zenith on point-sources. However, comparing the observed photometry to the photometry computed from the spectra and normalized in the $i-$band, we observed that the spectroscopic photometry was lacking $\sim 40\%$ of the flux in the \textit{u}-band. This apparent lack of flux was traced back to three different processes: extinction from the Milky Way galaxy, atmospheric absorption, and atmospheric refraction. The galactic extinction is due to the presence of dust in the Milky-Way. This was corrected applying the $E(B$-$V)$ maps of Schlegel et al. (1998). This correction is small ($\sim7$\% at 4000\AA~) but necessary as we compare the spectrophotometry of VIMOS spectra with broad band photometry corrected from the Milky Way extinction. The atmospheric absorption depends on the quantity of airmass on the light path. To correct for this effect we used the prescription of Patat et al. (2011) which is defined for the Paranal observatory. Finally, the atmosphere is equivalent to a small angle prism before entering the telescope and this leads to a spread of the incoming light into a small spectrum with length depending on the airmass and parallactic angle, the angle of the slit to the zenithal angle. As this small pseudo-spectrum is produced before entering the spectrograph slit, a significant fraction of the flux may be lost when entering the 1 arcsecond slits used for the VUDS observations, the effect being negligible beyond $\sim4500$\AA ~observed wavelength and becoming increasingly stronger going further down in wavelength towards the UV domain. This loss was estimated and then corrected using a geometrical model for each source based on its light moments (size, ellipticity) to compute which fraction of the flux was out of the VIMOS slit.  
With these corrections implemented, the magnitudes measured from the \textit{u}-band imaging agree to within 0.02$\pm$0.3 magnitude with the \textit{u}- band magnitudes as derived from the spectra, the standard deviation being dominated by \textit{u}-band photometric errors for \textit{u}-band magnitudes ranging up to u$\sim$27. 

We use the VUDS sample with redshift reliability flags 2, 3 and 4 (primary and secondary objects), the highest reliability flags as defined in Le F\`evre et al. (2014) with a confidence level of $\sim75$\%, 95\% and 100\% for the spectroscopic redshift measurements for these three flags respectively (see Le F\`evre et al. 2005a, 2013a, and 2014 for more details about the redshift flag system). 
In addition, we select only objects which have the highest signal to noise ratio (SNR) spectra. We compute the SNR from the mean and $1\sigma$ standard deviation of the spectra continuum in two regions 1345-1395\AA~and 1415-1515\AA~ rest-frame chosen to avoid emission or absorption features. We only consider galaxies for which the spectrum has a signal-to-noise ratio higher than $5$ per pixel at $z<4$ ($SNR>12$ per resolution element) and higher than 3.5 at $z>4$ ($>8.5$ per resolution element). 

The sample we use in this study contains 2127 galaxies with $2.5<z<5.5$, the largest sample of galaxies with high SNR spectra covering this redshift range.



\section{Measuring the IGM transmission  with spectral fitting}

\label{measure_fit}

\subsection{IGM transmission: definition}
\label{def_IGM}
The IGM transmission has been mainly studied from QSOs. As a short summary measuring this quantity from QSOs is performed fitting a power law on regions redward to the Ly$\alpha$ emission and free of emission or absorption features. The power law is then extrapolated blueward of the Ly$\alpha$ line. The transmission is then the result of the ratio between the power law and the QSO spectrum in the region 1070-1170\AA~.\\

Working with a galaxy sample we have opted to measure the Ly$\alpha$ transmission from a model fit of the observed spectra. The best fit spectrum is produced from a $\chi^2$ minimisation on models built combining stellar population synthesis models, dust extinction, and IGM templates. The IGM transmission is then measured directly on the IGM template that has been selected for the best fit (as described in the next Section). The transmission is computed shortward of the Ly$\alpha$ line in the wavelength range $1070 \leq \lambda \leq 1170$\AA ~as measured on the IGM template, this wavelength range avoiding immediate proximity effects of the circum-galactic medium (CGM) in the vicinity of the galaxy itself.

\subsection{Spectral fitting software : GOSSIP+}
\label{gossip}

GOSSIP (Galaxy Observed-Simulated SED Interactive Program) is a software created to fit the spectro-photometric flux of galaxies, including spectra, against a set of synthetic models (Franzetti et al., 2008), and was born in the framework of previous large galaxy surveys like the VVDS (Le F\`evre et al. 2005a). The aim is to find the model galaxy that best reproduces the observed data, using spectra, photometric magnitudes, or both jointly, making it a unique tool. The result of the fit can be used to estimate a number of physical parameters like the star formation rate (SFR), age, or stellar mass of the observed galaxy through the computation of their Probability Distribution Function (PDF). GOSSIP correlates spectra and Spectral Energy Distribution (SED) with model spectra covering a range of age, metallicity, dust extinction E(B$-$V), and star formation histories (SFH). The range of models used in this study is presented in Table \ref{Paramspace} and is discussed in Section \ref{results}. The model spectra are computed from state of the art galaxy population synthesis models: BC03 (Bruzual \& Charlot, 2003) and M05 (Maraston, 2005). An object SED and/or spectrum is cross-correlated against all model spectra and the best model is identified by means of the best reduced $\chi^{2}$. The new upgraded version of the software used in this paper, GOSSIP+, has been developed with new functionalities and will be described in details in a forthcoming paper (Thomas et al, in preparation). For the purpose of the present paper the major improvement included in GOSSIP is related to the treatment of the IGM. While most of the SED-fitting programs use the M95 prescription which gives, for a given redshift, a single IGM transmission curve as a function of wavelength, GOSSIP can choose at each redshift among seven different IGM curves with different IGM transmission, as presented in the next section. 
This allows to choose among different possible transmission curves, and hence enable to identify IGM values deviating from the mean transmission in case the IGM transmission is indeed varying.

\subsection{IGM transmission templates}
\label{igmpresc}

The flux we observe from distant galaxies is the result of two distinct set of processes: emission and absorption from the galaxy itself and its immediate surroundings in the CGM, and absorption from \ion{H}{I} clouds located between the observer and the galaxy. The latter is the result of the presence of gas clouds along the LOS that absorb the light blueward of the Ly${\alpha}$ line at $1216$\AA~. The models used to reproduce this effect exhibit several differences: M95 is based on an empirical model that uses a Poissonian distribution for the Ly$\alpha$ forest (with a fixed doppler parameter). M06 is based on cosmological simulations in $\Lambda$CDM from which the Ly$\alpha$ forest number density as well as the Doppler parameter are derived (Meiksin \& White 2004). Comparing these models to observations along the LOS of QSOs M95 predicts a transmission too low to reproduce the observed IGM attenuation, while M06 is more in agreement with observations (e.g. Inoue et al. 2014). In this paper, we base our spectral fitting on M06, augmented by the possibility of using a range of 7 different transmission templates at any given redshift, and we refer the reader to M06 for a detailed comparison between the M06 and M95 models.

Both M95 and M06 produce, for a given redshift, a single attenuation curve. This is equivalent to say that at a given redshift all lines of sight towards distant objects are populated by the same number of gaseous clouds with similar properties, producing the same transmission independently of the position of the observed galaxy or QSO in the sky. As already quoted by M95, this is not expected to be the case because of the way absorbing clouds are distributed along the LOS, but experimental measurements of the transmission dispersion are scarce in the literature and this point remains open as noted in the Introduction. It is also unclear whether the covering factor of clouds along the line of sight would affect the flux transmission in the same way for a point-source QSO or for an extended galaxy. Because of these uncertainties we have therefore allowed the IGM to vary in our SED/spectra fitting prescription to enable a measurement of the transmission scatter. This adds an extra free parameter in the SED or spectral fitting.

To test variations in the IGM transmission, we construct IGM templates in addition to the mean IGM of M06. The range of possible transmission is set using the $\pm1\sigma$ range of IGM transmission from M95 (see their figure 3.a at $z=3.5$). To better probe the range of possible IGM transmission, we added transmission templates computed from the $\pm 0.5 \sigma $ and the $\pm1.5\sigma$ around the mean. 
It is important to note that our IGM templates are defined to probe a large range of IGM transmission, independently of whether the $\pm1 \sigma$ model values of M95 are right or wrong. Our logic is to use templates which  reproduce reasonably well the wavelength dependence of the IGM transmission, whether the transmission is high or low (even if some of our templates might be un-physical when saturating to a transmission of one at $z\sim3$). The fitting process is then free to choose which of the IGM templates is best in reproducing each observed galaxy spectrum, in combination with all other model parameters. 

\begin{figure}[h!]
\centering
\includegraphics[width=9cm,height=10cm]{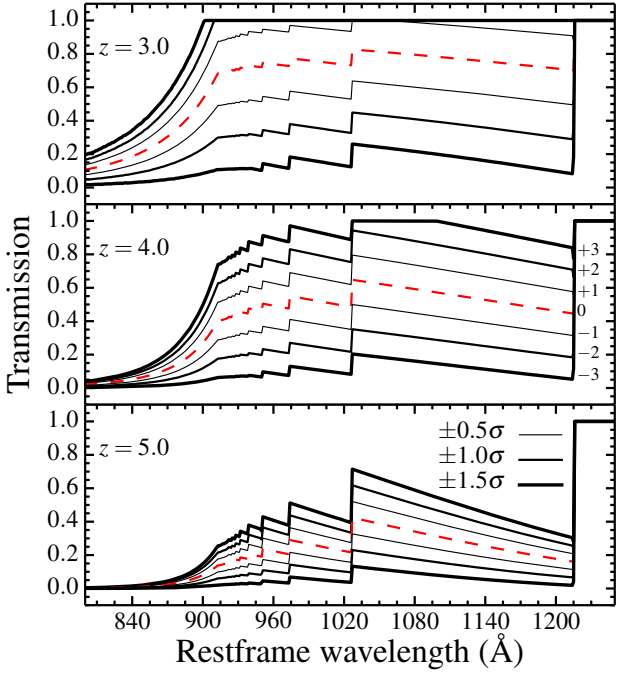}
\caption{The 7 IGM transmission templates used in this work at $z=3.0$, $z=4.0$ and $z=5.0$. The mean M06 IGM transmission is represented by the dashed red line. The $\pm 0.5\sigma$, $\pm1.0\sigma$ and $\pm1.5\sigma$ curves are the 6 black curves of different thickness. The thicker lines represent $\pm1.5\sigma$.}
\label{IGMvar}
\end{figure}

GOSSIP+ calculates the 7 IGM templates using the transformation described above for each galaxy depending on its redshift, and these templates are then used in the fitting process for that galaxy.
The set of templates allowed in our study at $z=3.0$, $z=4.0$ and $z=5.0$ (based on M06 models) are shown in Fig. \ref{IGMvar}. Each template is identified by an \textit{`identification number'} (hereafter $id$) $0$, $\pm1$, $\pm2$ and $\pm3$ (middle panel). The $id=0$ (red curve) is the M06 mean IGM transmission curve.  The $id$ ranges from $-3$ (smallest transmission; $-1.5\sigma$) to $+3$ (highest transmission; $+1.5\sigma$). The range of IGM transmission is large at any redshift and allows us to explore the dispersion around the mean. With our adopted templates the IGM transmission at 1100\AA ~may vary from 19\% to 100\% at z=3.0 and from 5\% to 50\% at z=5.0.

In summary the spectral fitting procedure with GOSSIP+ allows to consider very different lines of sight from one spectrum to another, as the fitting algorithm tests each galaxy model with all of the seven IGM transmission templates.

\subsection{Simulation of spectra: ability to recover  the IGM transmission}
\label{simu}
We performed extensive spectral fit simulations to test the ability of GOSSIP+ to retrieve the IGM transmission. To be representative of our galaxies we simulate 50 times each 3-tuple (redshift, $i_{AB}$, SNR) that we have in our sample of 2127 galaxies. In our data the redshift ranges from 2.5 to 5.5, the SNR per resolution element ranges from 5 to 25 (at z$<$4) and from 3.5 to 10 (at z$>$4), the selection magnitude $i_{AB}$ goes from 22 to 25. The final sample of mock galaxies is then populated by almost 108000 objects.
Each mock galaxy is based on a galaxy model randomly chosen from a BC03 model library composed from a wide range of model parameters. The E(B$-$V) extinction ranges from $0$ to $0.5$. The metallicity is chosen in the range Z=[0.004, 0.05]. The star formation history is a delayed exponential (see section \ref{results}) with a timescale that ranges from $0.1$ to $1.0$ Gyr. Ages are limited by the age of the Universe at the given redshift and for the selected Cosmology. Each model spectrum is multiplied by one of the seven IGM templates, selected randomly. Poisson noise is then added to the data as the square root of the simulated flux.

The synthetic galaxies are then fitted with GOSSIP+. As for the data we ignore emission lines regions for the fit (Ly$\alpha$). Since we created those mock galaxies with BC03 templates we chose to fit the model spectra using a library computed from the M05 synthesis models. This library is computed with an exponentially delayed SFH with a timescale that ranges from $0.1$ to $1.0$ Gyr (with a step of 0.1 Gyr), E(B$-$V) can take values in [0.0, 0.05, 0.1, 0.2, 0.3, 0.4, 0.5] and the metallicity ranges from Z=0.001 to Z=0.04. This library is computed with a Chabrier IMF.

\begin{figure}[h!]
\includegraphics[width=9cm,height=8cm]{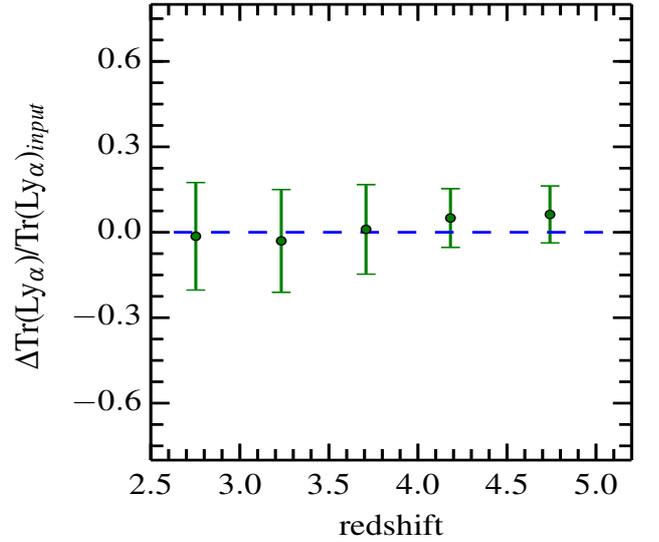}
\caption{Evolution with redshift of the difference in transmission $\Delta$Tr(Ly${\alpha}$)/Tr(Ly${\alpha}$) between the input simulated spectra and the result of the fit with GOSSIP+.  }
\label{IGMsimu2}
\end{figure}

The relation between the simulated transmission and the transmission obtained by the fitting process is given by $\Delta$Tr(Ly${\alpha}$)/Tr(Ly$\alpha$)$_{input}$ defined as:

\begin{equation}
\frac{\Delta Tr(Ly\alpha)}{Tr(Ly\alpha)_{input}}= (Tr(Ly\alpha)_{input}-Tr(Ly\alpha)_{output})/Tr(Ly\alpha)_{input}
\end{equation}
The results of the simulation on this quantity is presented in Figure \ref{IGMsimu2}, representing the accuracy of GOSSIP+ in retrieving the correct Tr$(Ly_{\alpha})$ as a function of redshift for galaxies with a similar range of properties as in the VUDS sample.
For the whole simulated sample $\Delta Tr(Ly_{\alpha})/Tr(Ly\alpha)_{input}$=$-0.02\pm0.14$. The $\Delta \mathrm{Tr(Ly}{\alpha})/Tr(Ly\alpha)_{input}$ ranges from -0.01$\pm$0.18 at z=2.7 to 0.06$\pm$0.10 at z=4.8. There is therefore very little deviation from the mean and this shows that GOSSIP+ can retrieve the correct Ly${\alpha}$ transmission at all redshifts and over the spectra SNR considered in this study. We note that the standard deviation in our simulations is slightly better at higher redshifts, likely the result of a larger Ly$\alpha$ dropout break at the higher redshifts and stronger IGM features, easier to retrieve in the fitting process. 
GOSSIP+ is able to retrieve the correct IGM transmission in $\sim$74\% of the fits. There are 26\% of the cases when GOSSIP+ finds another set of parameters that reproduces better the simulated spectra, mostly resulting from the degeneracy between dust and IGM extinction at lower SNR as discussed in Section \ref{dust}. 
The influence of this IGM prescription on other physical parameters when fitting spectra/SED will be fully discussed in a forthcoming paper (Thomas et al, in prep).

\section{Evolution of the mean IGM transmission and its dispersion as a function of redshift}
\label{results}

Here we present the IGM transmission obtained from the fit of VUDS galaxy spectra with redshifts between $z=2.5$ and $z=5.5$, with the selection as described in Section \ref{data} and using GOSSIP+ as discussed in Section \ref{gossip}.  
Since we combine the IGM transmission templates with different galaxy models we expect the results to be as much independent from stellar population synthesis models as possible. We run GOSSIP+ twice, using both a BC03 model library and a library computed with M05 models. The large parameter space explored during the fitting process is summarized in Table \ref{Paramspace}.
 
\begin{table}
\caption{Parameter space used for the spectral fitting, using existing BC03 and M05 stellar population models. }
\label{Paramspace}    
\centering                       
\begin{tabular}{c c c }       
\hline\hline                
Parameter & BC03 & M05 \\   
\hline                       
IMF & Chabrier & Chabrier \\ 
 $\tau_{SFR}$ in Gyr & \multicolumn{2}{c}{0.1 to 1.0, 0.1 steps}  \\ 
Metallicity [Z] &  0.004,0.008,0.02,0.05 & 0.001, 0.01, 0.02, 0.04 \\ 
 E(B$-$V) & \multicolumn{2}{c}{0, 0.1, 0.2, 0.3, 0.4, 0.5} \\ 
 E(B$-$V) at $z>4$ & \multicolumn{2}{c}{0, 0.05, 0.1, 0.15} \\ 
Ages (Gyr) & \multicolumn{2}{c}{0.05 to 4.0}  \\ 
IGM & \multicolumn{2}{c}{7 curves} \\ 
\hline                                  
\end{tabular}
\end{table}

The star formation history used in this study is a delayed exponential. It is defined as:
\begin{equation}
SFR(t)=\frac{1}{\tau^{2}} \times t \times \exp\frac{-t}{\tau}
\end{equation}
where $\tau$ is the SFH timescale. It corresponds to the time after which the SFR is maximal. A range of $\tau$ from 0.1 to 1.0 Gyr allows us to produce very different SFHs, from a fast rising star formation rate to an early peak in star formation followed by a rapid decrease, to a more extended period of almost continuous star formation. We use the initial mass function (IMF) of Chabrier (2003). The dust extinction is applied through the value of E(B$-$V) using the Calzetti et al (2000) law. As our model spectra do not include emission lines, the wavelength domain around known emission lines in the observed spectra are therefore ignored in the fitting process.

\begin{figure*}
\resizebox{\hsize}{!} {\includegraphics[width=8cm,height=9cm,clip]{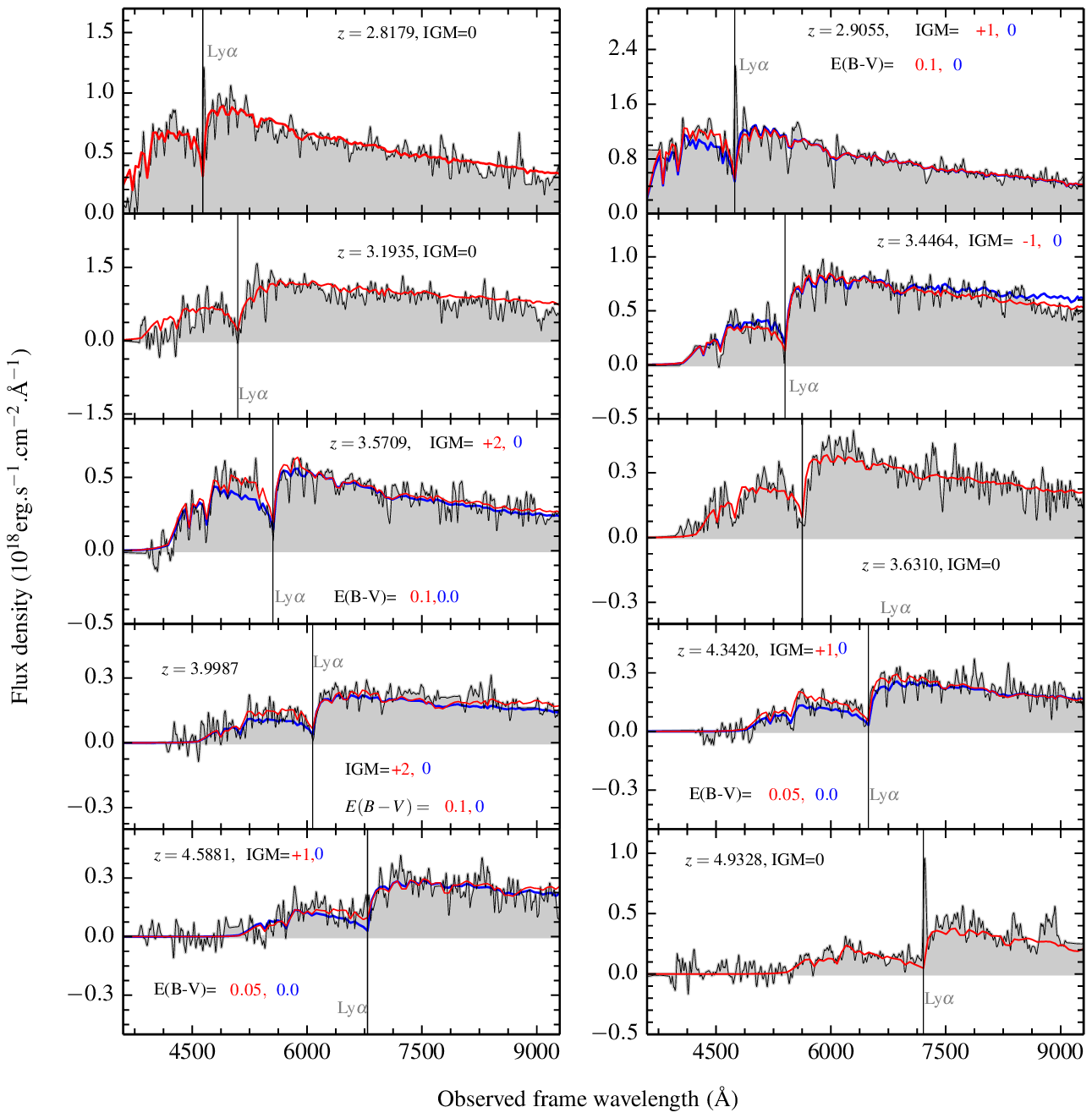}}
\caption{Example of spectral fitting for 10 representative galaxies in the VUDS sample, using the BC03 model spectra. The IGM `id' (cf section 2.) and the redshift of the galaxy are indicated for each fit. The best fit obtained with our prescription varying the IGM transmission is in red, while the best fit with the IGM transmission set to the mean M06 value is shown in blue for comparison (the E(B$-$V) dust extinction is reported for both). For 4 objects the best fit finds an IGM equal to the mean, while for the 6 others GOSSIP+ selects an IGM template different from the mean}
\label{exemplefit}
\end{figure*}

Figure \ref{exemplefit} presents the results of spectral fitting for ten representative objects at different redshifts between z=2.5 and z=5.5. For four of them, the fit locks-in to an IGM identical to the mean IGM. For the other 6 GOSSIP+ identifies an IGM transmission different from the mean.

\begin{figure}[h!]
\centering
\includegraphics[width=9cm,height=10cm]{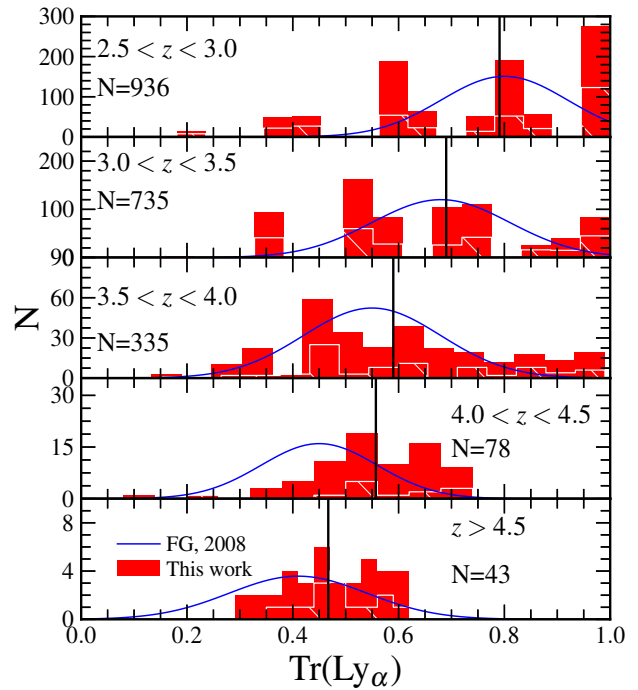}
\caption{IGM transmission Tr(Ly${\alpha}$) measured from 2127 VUDS galaxies in increasing redshift bins covering from z=2.5 to z=5.5. The black vertical lines represent the median of the distributions in each redshift bin. A large spread in transmission is observed at all redshifts. The red shaded histograms include all galaxies with flag 2,3,4, while the white striped histograms represent flag 2 only. No significant difference is found between the IGM properties of flag 3 and 4 and flag 2. The distribution of IGM transmission is somewhat discretized because of the 7 IGM templates used in the spectra fitting. The median and dispersion values measured by Faucher-Gigu\`ere et al. (2008b) using QSOs are represented by the blue distribution in each redshift bin. The median values from VUDS galaxies are in excellent agreement with Faucher-Gigu\`ere et al. (2008b) for z$<$4, while the VUDS results seem to show higher transmission at z$>$4 (see text for a discussion). The dispersion values are also in excellent agreement at all redshifts. }
\label{IGMlyatrbins}
\end{figure}

\begin{figure}[h!]
\centering
\includegraphics[width=9cm,height=8cm]{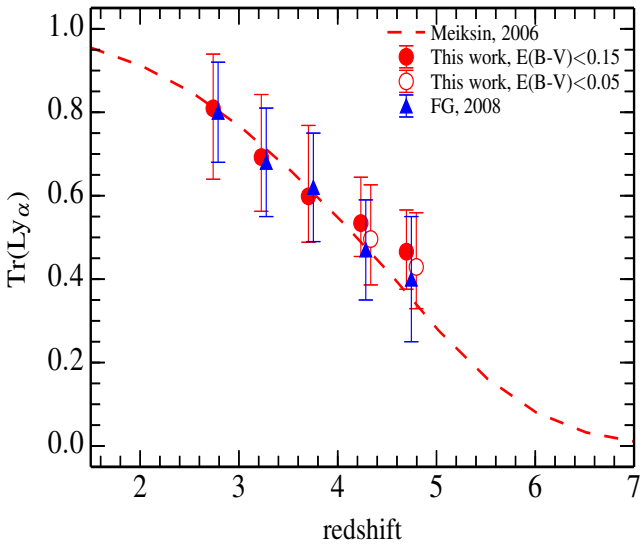}
\caption{Median IGM transmission Tr(Ly$\alpha$) derived from the line of sight towards 2127 VUDS galaxies (full red circles). IGM transmission values are computed from the distribution of IGM transmission measurements resulting from the fit with GOSSIP+, in redshift bins $\delta$z$=$0.5 at a mean redshifts $2.75$, $3.25$, $3.75$, $4.25$ and in a redshift bin $\delta$z=0.8 at z=$4.8$, and using a dust extinction limit E(B-V)$\leq$0.15. The error bars represent the 68\% interval in the distribution of IGM transmission values as measured from the 2127 LOS probed by the VUDS galaxies. The empty red circles at z$>$4 represent the measurements obtained when setting the dust extinction limit to E(B-V)$\leq$0.05. The blue triangles and associated error bars are median transmission and 1$\sigma$ dispersion values taken from Faucher-Gigu\`ere et al. (2008b). At $z>4$, points have been artificially shifted in redshift for clarity.}
\label{IGMmean}
\end{figure}

We perform the fit with GOSSIP+ of all the 2127 galaxies in the VUDS sample described in Section \ref{data}. 
Measurements of the median IGM transmission  are presented in Table \ref{result_table} and in figures \ref{IGMlyatrbins} and \ref{IGMmean}. The dispersion around the median is computed as the value including 68\% of the full sample in each redshift bin. 
Figure \ref{IGMlyatrbins} shows the distribution of Tr(Ly$_{\alpha}$) in $5$ redshift bins covering from z=2.5 to z=5.5. Tr(Ly$_{\alpha}$) is computed on the IGM templates obtained from the GOSSIP+ fit for each VUDS galaxy. As shown in Figure \ref{IGMlyatrbins} there is no significant difference between the distribution of transmission observed for the most reliable flag 3 and 4 in our sample compared to the lower reliability flag 2.
We note that the use of the BC03 or M05 templates in the fitting process provides undistinguishable results on measured IGM properties at any redshift. This effect is expected because the main difference between BC03 and M05 is the treatment of TP-AGB stars that have an influence in the red part of the model spectra. The median and mean transmission values are nearly identical in our sample. In the following we use indistinctly {\it mean transmission} or {\it median transmission}.
A large range of IGM transmission is observed at all redshifts.

\begin{table*}
\centering                       
\caption{Median IGM transmission and dispersion around the median as observed from the VUDS spectra fitting. The dispersion is computed as 68\% of the distribution. The transmission values are indicated for the GOSSIP+ fits performed using the BC03 and M05 composite stellar population models. Comparing these results show that IGM transmission and dispersion values do not depend on the stellar population used in the fit. At z$>$4 we indicate two different measurements, one is obtained when the dust extinction in galaxies is let free to vary up to E(B$-$V)=0.15 while the other is for dust constrained to  E(B$-$V)$\leq$0.05 (see Section \ref{dust}) . Results from the VUDS galaxies LOS are compared to the mean IGM transmission from Becker et al. (2013) and to the IGM dispersion measured by Faucher-Gigu\`ere et al. (2008b).}
\label{result_table}    
\begin{tabular}{c c c c c c c c c c }       
\hline\hline                
         &          & \multicolumn{6}{c}{VUDS galaxies}                                         & \multicolumn{2}{c}{QSOs}  \\ \hline
         &          & \multicolumn{4}{c}{Median IGM Transmission Tr(Ly$\alpha$)} & \multicolumn{2}{c}{Dispersion}   & IGM transmission    & IGM dispersion  \\
Redshift & Median z & \multicolumn{2}{c}{Dust E(B$-$V)$\leq$0.15}  & \multicolumn{2}{c}{Dust E(B$-$V)$\leq$0.05} & \multicolumn{2}{c}{$\pm$68\%}   &  (Becker et al. 2013) & Faucher-Gigu\`ere et al. (2008b) \\
         &          & BC03  & with M05  &   BC03  &  M05 &  BC03  &  M05 & \\ \hline                       
$2.5<z<3.0$ & 2.75 & 0.79$_{-0.005}^{+0.004}$ & 0.78$_{-0.006}^{+0.004}$  & - & - & $_{-0.17}^{+0.13}$ & $_{-0.18}^{+0.13}$ & 0.80 & 0.12 \\    %
$3.0<z<3.5$ & 3.22 & 0.69$_{-0.007}^{+0.005}$ & 0.67$_{-0.006}^{+0.007}$  & - & - & $_{-0.13}^{+0.17}$ & $_{-0.14}^{+0.18}$ & 0.70 & 0.13 \\    %
$3.5<z<4.0$ & 3.70 & 0.59$_{-0.006}^{+0.009}$ & 0.59$_{-0.007}^{+0.009}$  & - & - & $_{-0.11}^{+0.15}$ & $_{-0.12}^{+0.15}$ & 0.59 & 0.13 \\    %
$4.0<z<4.5$ & 4.23 & 0.55$_{-0.010}^{+0.012}$ & 0.56$_{-0.012}^{+0.015}$  & 0.49$_{-0.014}^{+0.015}$ & 0.50$_{-0.012}^{+0.014}$ & $_{-0.08}^{+0.11}$ & $_{-0.11}^{+0.13}$ & 0.47 & 0.12 \\  
$z>4.5$     & 4.77 & 0.46$_{-0.012}^{+0.020}$ & 0.45$_{-0.016}^{+0.020}$  & 0.42$_{-0.019}^{+0.020}$ & 0.42$_{-0.016}^{+0.020}$ & $_{-0.09}^{+0.10}$ & $_{-0.10}^{+0.13}$ & 0.35 & 0.13 \\   
\hline                                  
\end{tabular}
\end{table*}

As expected the mean IGM transmission decreases with redshift (Fig. \ref{IGMmean}), the first time this is clearly demonstrated from galaxy spectra. The  IGM transmission derived from VUDS galaxy spectra is in remarkable agreement with the mean transmission in the IGM model of M06 up to z=4.  At a redshift z$>$4 we find tentative evidence that the median IGM transmission is slightly higher than expected from the M06 model. The difference in transmission between the observed and theoretical model is $\Delta$Tr$(Ly_{\alpha})$=$0.03$ at 4.0$<z<$4.5 and reaches $\Delta$Tr$(Ly_{\alpha})$=$0.09$ at 4.5$<$z$<$5.5 (corresponding to $\sim$18\% of the theoretical value). Whether this is a real physical result, or the result of degeneracy between IGM transmission and dust internal to the galaxies is further discussed in Section \ref{dust}. We report in Figure \ref{IGMmean} IGM transmission values obtained for two different dust extinction conditions with dust extinction limited to E(B$-$V)$\leq0.15$ and E(B$-$V)$\leq$0.05 in the fitting process. 
 
The other important result from our study is the high dispersion of Tr(Ly$_{\alpha}$) observed for all redshifts in the range we have explored. To compute the dispersion in IGM transmission, we correct in quadrature the observed dispersion for the dispersion resulting from our fitting method as derived from the simulations presented in Section \ref{simu} for spectra with SNR in the range of VUDS spectra.
The resulting transmission dispersion is found to be large with a slow decrease with increasing redshift: it ranges from $\sigma\sim0.15$ 
at redshift $2.8$ to $\sigma\sim0.10$ at redshift $4.8$. We stress that these are effective $1\sigma$ values, and we observe that all 7 IGM templates covering the range $\pm1.5 \sigma$ allowed in our fitting process are used at any redshift. We find that in 72\% (fitting with BC03 templates) and 75\% (with M05) of the cases GOSSIP+ has chosen one of the 6 additional transmission curves built besides the mean model value. The observed variation in Tr(Ly${\alpha}$) is large enough to make it necessary to consider a large range of IGM variation when fitting the spectra or SED of galaxies as well as in selecting galaxies based on colour-colour selection (see Section \ref{selec}).

IGM transmission studies from QSOs often quote the error on the mean $\sigma(Tr(Ly\alpha))$=$\sigma(Tr(Ly\alpha))_i/\sqrt(N)$ where N is the number of independent measurements. Faucher-Gigu\`ere et al. (2008a) find $\sigma(Tr(Ly\alpha))\sim0.008$\% at $z=3$ and $\sim1$\% at z=4 in redshift bins $\delta$z=0.2, while Becker et al. (2013) quote about 1\% in $\delta$z=0.1 bins.
Only few studies have reported the dispersion around the mean IGM transmission. Faucher-Gigu\`ere et al. (2008b) present the distribution in transmission from z=2.2 to z=4.6 and derive a dispersion ranging from $\sigma=0.11$ to 0.15 in in $\delta$z=0.2 bins.  Songaila (2004) computed the IGM transmission from 50 high redshift quasars and its evolution with redshift. From their data they give a transmitted fraction ranging from 46\% to 90\%between z=2.5 and z=3.5 , and from 20\% to 80\% between z=3.5 and z=4.5, a range similar to Faucher-Gigu\`ere et al. (2008b) and our VUDS measurements. 

The dispersion derived by Faucher-Gigu\`ere et al. (2008b) is based on breaking-up the part of the same QSO spectrum affected by the IGM  in segments of 3 h$^{-1}$Mpc along the LOS and measuring the dispersion among these from different LOS to emulate a large number of independent measurements. The dispersion measurements on VUDS galaxies are done from the wavelength domain 1070$-$1170\AA, corresponding to a scale of about 70 h$^{-1}$Mpc at z$\sim$3.5. As the sizes of IGM absorbers are thought to have remained roughly constant with time and less than 100 h$^{-1}$kpc (Cen 2012), much smaller than 3  h$^{-1}$Mpc, our measurements should be directly comparable to those of Faucher-Gigu\`ere et al. (2008b). To compare dispersion values we also need to use the same redshift bins as part of the dispersion is coming from the evolution in the transmission itself, changing by 5-10\% over $\delta z$=0.5 for the redshift range considered. Using the $\delta z$=0.2 values from Faucher-Gigu\`ere et al. (2008b) we find that the dispersion from their data in 
$\delta z$=0.5 is increased by about 20\% which brings our dispersion measurements in even better agreement. Although working with smaller samples, the reverse is also verified when computing the dispersion in $\delta z$=0.2 bins we find a dispersion similar to that of Faucher-Gigu\`ere et al. (2008b). We therefore conclude that the dispersion observed in our data on galaxies is in excellent agreement with the dispersion found along the LOS of QSOs and reflects the intrinsic properties of the IGM. 

\begin{figure*}
\centering
\includegraphics[width=15cm]{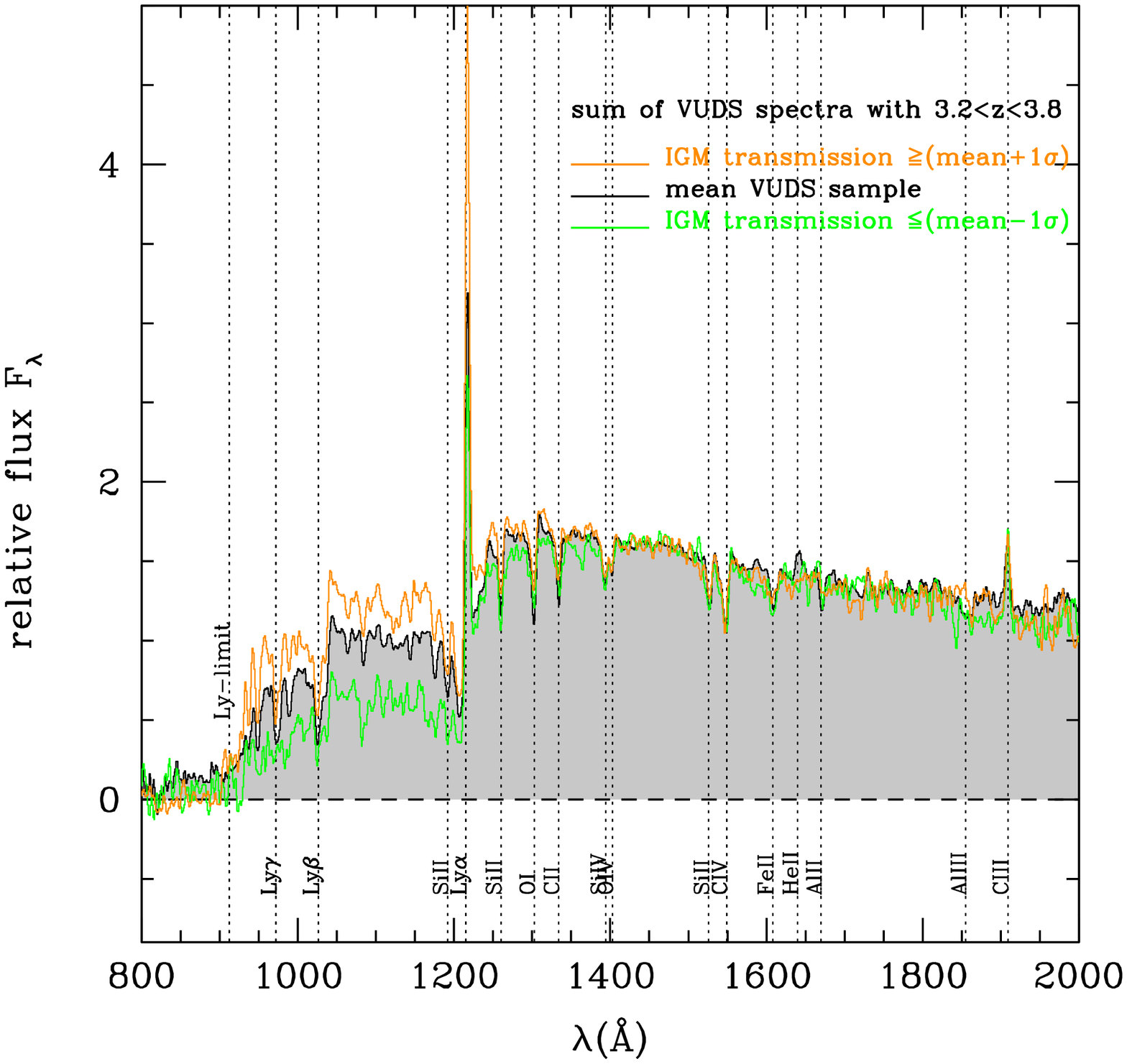}
\caption{Average (stacked) spectra for VUDS galaxies with $3.2 \leq z \leq 3.8$ (z$_{med}$=3.4) for which GOSSIP+ identifies an IGM transmission $\geq1\sigma$ above (gold spectrum) and below (green spectrum) the mean IGM transmission of M06. 
The average VUDS spectrum over all galaxies in the sample is shown as the black shaded-grey spectrum. All three spectra are normalized 
in the range 1400$<$$\lambda$$<$1510\AA.} 
\label{stacks}
\end{figure*}

We plot in Figure \ref{stacks} the average of 60 spectra with the maximum IGM transmission $Tr(Ly\alpha) \geq Tr(Ly\alpha)_{mean} + 1 \sigma$ and for 53 spectra with the minimum IGM transmission lower than the mean $-1 \sigma$, compared to the mean spectrum, for galaxies with 3.2$\leq$z$\leq$3.8 (and with the highest reliability flags 3 and 4). 
The spectra are normalized to the same continuum value redward of Ly$\alpha$ (in the range free of strong lines between 1345--1395
 and 1415--1515\AA).
The difference observed between galaxies with IGM transmission measured at $\pm 1 \sigma$ from the mean and the mean spectrum of the full sample is quite striking and provides further support to the observed range in IGM transmission reported above.
The stacked spectrum of the full sample and spectra at $\pm 1 \sigma$ from the mean are identical, within errors, in slope or in absorption lines strengths. This shows that stellar populations at the source emission are very similar, lending additional support that the IGM and not the source properties are responsible for the observed spectra differences below Ly$\alpha$. The only notable difference is the strength of the Ly$\alpha$ emission line, with equivalent width EW(Ly$\alpha$)=16\AA ~for the high IGM transmission spectra and  EW(Ly$\alpha$)=10\AA ~for low transmission spectra. This difference might further indicate that Ly$\alpha$ photons are more absorbed by the CGM of the galaxy in this latter case. 

\section{Evolution of the Ly$\alpha$ effective optical depth}
\label{teff}

\begin{figure*}[h!]
\resizebox{\hsize}{!} {\includegraphics[width=7cm,height=4cm,clip]{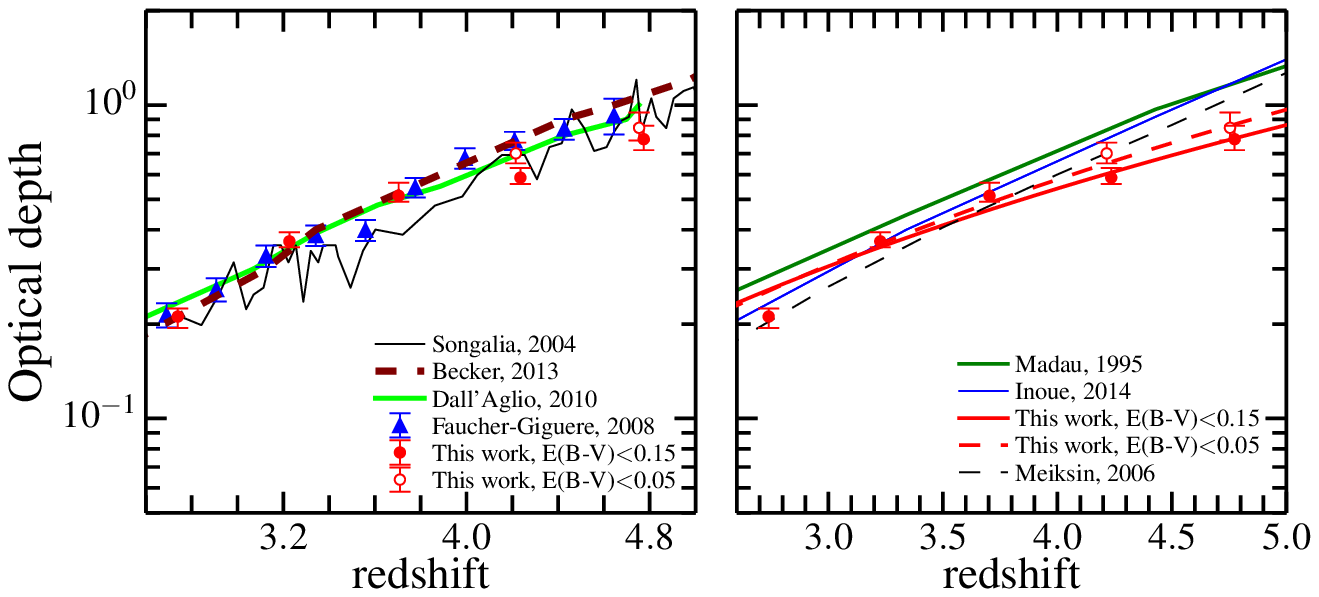}}
\caption{(Left panel). The HI optical depth $\tau_{eff}$ as computed from the measured IGM transmission in the VUDS galaxy data (red filled circles for E(B-V)$\leq$0.15 and empty red circles for E(B-V)$\leq$0.05). The error bars on  $\tau_{eff}$ represent the error on the mean (defined as $\sigma(Tr(Ly\alpha))/\sqrt(N)$) on the 2127 LOS probed by the VUDS galaxies. 
The data of Songaila (2004) are indicated by the black line and the brown line represents the measurements from Becker (2013). The blue triangles are from Faucher-Gigu\`ere et al. (2008a) and the green line indicates the measurement of Dall'Aglio et al. (2008). 
(Right panel). Power law fit of the VUDS HI optical depth $\tau_{eff}$ as a function of redshift, with $\tau_{eff}=C (1+z)^{\gamma}$ with $\gamma=2.55$ and $C=0.0089$ for the points computed with E(B-V)$<$0.15 and $\gamma=2.81$ and $C=0.0062$ for the point computed with E(B-V)$<$0.05. This is compared to the M06 model (black dashed line), M95 models (green) and Inoue et al, 2014 (blue).}
\label{tau}
\end{figure*}

The mean IGM transmission is related to the HI effective optical depth, $\tau_{eff}(z)$, by 
\begin{equation}
\tau_{eff} \equiv - \ln Tr(Ly_{\alpha}) \equiv - \ln(e^{-\tau_{HI}}).
\end{equation}
We use our data to obtain the evolution of $\tau_{eff}$ with redshift as presented in Figure \ref{tau}. This is compared with $\tau_{eff}$ derived from QSO spectra as there is no measurement derived from galaxy spectra in the literature. Our measurements from z=2.5 to z=4 are in excellent agreement with the mean $\tau_{eff}$ reported in the literature from samples with large numbers of QSOs, as shown in the left panel of Figure \ref{tau}, e.g. the measurements of Becker et al. (2013) as well as to the data of Faucher-Gigu\`eres et al. (2008a) and Dall'Aglio et al. (2008). 
This agreement is quite remarkable given the fact that we are using a sample of faint galaxies to derive the optical depth. This result therefore gives further merit to the spectral fitting procedure that we have developed letting the IGM free to vary. 

Our highest redshift measurements at z=4.2 and z=4.75 present a lower optical depth with respect to M06, the result from the higher transmission reported in Section \ref{results}. At these redshifts the IGM transmission values are somewhat degenerate with the dust extinction in galaxies as discussed in Section \ref{dust}. We therefore present the IGM transmission values obtained in the case when the E(B$-$V) extinction at the source is let free to vary up to E(B$-$V)=0.15, as well as transmission values obtained for  lower dust extinction E(B$-$V)$\leq$0.05. In this latter case, the IGM transmission is significantly smaller and becomes compatible with QSO measurements and models within measurement errors as shown in Figure \ref{tau}. This is further discussed in Section \ref{dust}.

We fit our data with a power law of the form $\tau_{eff}= C \times (1+z)^{\gamma}$ which is generally used in the literature to reproduce the evolution of $\tau_{eff}$. The fit is presented in the right panel of Figure \ref{tau}. Our mean optical depth rises following a power law with $\gamma=2.55_{-0.07}^{+0.08}$ and $C=0.0089_{-0.0003}^{+0.0005}$.  
We also display as comparison the models from M06, M95 and Inoue et al (2014).

\section{Uncertainties in observations and models}
\label{discuss}

\subsection{Possible observational biases at z$>$4}

While the IGM transmission that we find in 2.5$<$z$<$4.0 is in excellent agreement with QSO measurements and the latest models from M06 and Inoue et al. (2014), we need to understand if the higher transmission values we find at z$>$4 are physical measurement or the result of some uncertainty or bias in our data or methods.
 
As presented in section \ref{data} the selection of VUDS galaxies was made mainly using their photometric redshift. Photometric redshifts were computed with the code LePhare (Ilbert et al. 2006, 2009) that uses the M95 prescription with a single mean IGM transmission at a given redshift. Therefore this selection should preferably select galaxies with the M95 IGM transmission. As the M95 transmission is somewhat lower than is observed from QSOs or the more recent simulations of M06 and Inoue et al. (2014) the photometric selection with Le Phare and the M95 IGM transmission would result in a sample bias, if any, towards galaxies with a lower transmission on their LOS than observed from QSOs. Our highest redshift data points at z=4.2 and z=4.8 seem rather to indicate a higher transmission, a trend opposite to what would be expected from our photometric redshift selection. If any such bias would exist, correcting it would further increase the difference between the observed mean and the model M06 mean, and it is therefore likely that our data is not affected by this type of bias. 

As described in Section \ref{data}, the VIMOS spectra are corrected for atmospheric extinction and refraction (in addition to galactic extinction negligible for the effects discussed here). The maximum flux correction applied to spectra is of $\sim$60\% at 3800\AA, while it is less than 25\% at 4500\AA ~and less than 10\% at 5500\AA. The mean distribution of the corrected spectroscopic fluxes agrees with the fluxes obtained from the multi-band photometry within 2\% (Section \ref{data}), with a $1\sigma$ standard deviation less than 5\% in the u--band once corrected from the magnitude errors for faint magnitudes $u<25$ in this band, and further confirmed from the brighter g--band magnitudes where $1\sigma<2$\%. These values are significantly lower than the IGM transmission dispersion of $\sim15$\% observed in our data. Furthermore, spectra flux calibration corrections affect mainly the u--band and would therefore affect the IGM transmission measurements for a redshift z$\sim$2.5, while residual errors in g--band corrections would affect measurements for z$<$3.0. These small residual instrumental calibration errors are unlikely to have an effect on the measured IGM transmission dispersion. 

Another possible observational bias is that the spectra of galaxies with the lowest IGM transmission are the most likely to be in our high SNR sample because the break produced in the spectrum is easiest to identify when measuring the spectroscopic redshifts of these distant galaxies. This would create a distribution skewed towards low transmission systems, which is again the opposite to what is observed in our data at z$>$4. We note that we use the LOS towards 43 high SNR galaxies at $z\sim4.8$, therefore still small and subject to cosmic variance. More LOS will need to be observed to confirm the trend for higher transmission observed compared to models at $z>4$.

It is also important to note that the IGM transmission in our sample is computed from extended galaxies rather than from point-like QSOs. This is further discussed in Section \ref{mod}.


\subsection{Dust versus IGM at z$>$4}
\label{dust}

The other main parameter that impacts the fit in the UV part of the spectra besides IGM transmission is dust extinction internal to the galaxies. A possible way to compensate a change in the IGM transmission is to change the dust content of the galaxy: if the IGM transmission is higher (smaller) the dust extinction would need to be higher (smaller) in the best fit. 

At z$<$4 we compute the E(B-V) values from the spectra fitting using a limit of E(B-V)=0.5 and obtain a mean value of E(B$-$V)=0.12 at $z\sim3$. This is in agreement with Cucciati et al. (2012) who found E(B$-$V)=0.14 at $z=3.0$ and Shapley et al. (2003) reporting E(B-V) in the range 0--0.2.  
At $z>4$, we computed our E(B-V) with a  limit of E(B-V)=0.15, as presented in section \ref{results}. We find a mean E(B-V) of 0.12 for these galaxies. At z$\sim$4 Tresse et al. (2007) and Cucciati et al. (2012) report a mean E(B$-$V)$\sim0.05-0.1$, somewhat lower than Ouchi et al. (2004) who computed E(B$-$V)$\sim$0.15 at z=4.7. At even higher redshifts Bouwens et al. (2013) report values E(B$-$V)$<$0.02-0.03. The E(B-V) is not well constrained from the fit of a UV-rest spectrum, so we use the E(B-V) computed from a full SED fitting of all photometric data points and we find that the mean E(B-V) at z$>$4 is E(B-V)=0.03. 

To study the dependency of dust extinction on our results we ran GOSSIP+ again for $z>4$ galaxies with three different limits: E(B$-$V)$_{max}$=0.05, E(B$-$V)$_{max}$=0.15 and E(B$-$V)$_{max}$=0.5. When the fit is able to explore a larger range of E(B$-$V) the mean E(B$-$V) increases. This results on a higher IGM transmission and therefore a lower optical depth. At z=4.2, The IGM tranmission goes from 49\% when E(B$-$V)$\leq$0.05 to 65\% when E(B$-$V)$\leq$0.5 at z=4.2. At z=4.8 the change is less pronounced and goes from 42\% for E(B$-$V)$\leq$0.05 to 51\% for E(B$-$V)$\leq$0.53. We also computed the mean E(B-V) values obtained when changing the high limit of E(B-V) at z$>$4. When the E(B$-$V) values are able to reach E(B$-$V)$_{max}$=0.5 the mean E(B$-$V) is 0.12 while it goes down to 0.03 when E(B$-$V)$_{max}$=0.05.

 Consequently, leaving the dust extinction limit E(B$-$V)=0.5 drives $\tau_{eff}$ away from the simulation of M06 as well as from the measurements derived from QSOs, while lowering the maximum E(B$-$V) allowed in the fit brings it closer. This is indirect evidence that the measured dust extinction in our galaxies might be lower than a straight fit leaving E(B$-$V) unconstrained would indicate. We therefore provide as our best measurement at z$>$4 the IGM transmission value obtained for a dust content of galaxies up to E(B$-$V)=0.15, and we also report the IGM transmission for a dust content limited to E(B$-$V)=0.05 (Table \ref{result_table}).

Another a priori hypothesis in our analysis is the dust extinction law. While we use Calzetti et al. (2000), the use of a different extinction law like that of the SMC (Pr\'evot et al. 1984) has been advocated for galaxies with lower metallicity at high redshifts (e.g. Speagle et al. 2014). 
In a preliminary analysis we used the SMC extinction law for the fit of our $z>4$ galaxies for E(B$-$V)$_{max}$=0.15. 
We find that the IGM transmission is even more driven away from the mean transmission from QSOs or models. Since the extinction law of the SMC has a steeper slope than the Calzetti (2000) law, the same E(B-V) value leads to a more pronounced extinction with the SMC extinction law. Thus, to compensate this effect the IGM transmission must be higher.

\begin{figure}[h!]
\centering
\includegraphics[width=9cm,height=8cm]{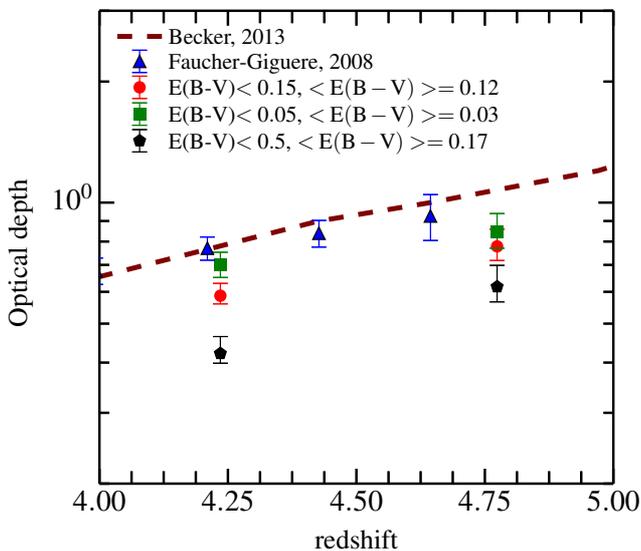}
\caption{Impact of dust extinction on optical depth measurements at $z>4$. The optical depth $\tau_{eff}$ is shown for different limits in E(B$-$V): E(B$-$V)$_{max}=0.05$ (in green), E(B$-$V)$_{max}=0.15$ (in red) and E(B$-$V)$_{max}=0.5$ (in black). The blue points are the measurement from Faucher-Gigu\`eres (2008b) and the brown line is from Becker et al. (2013). The redshifts for VUDS points with different E(B$-$V) at z=4.2 and z=4.75 have been artificially shifted for clarity.}
\label{comparedust}
\end{figure}

\subsection{Uncertainties in IGM transmission models, and angular projection effects}
\label{mod}

One important aspect to consider in comparing to models is that the dominant contribution to the continuum opacity is from a small number of absorbers at the largest column densities in a given LOS (Madau et al. 1996). As pointed out by Bershady et al. (1999), small number statistics on those absorbers
may cause large variations in the observed colours of high-redshift galaxies of the same intrinsic spectral type.
Moreover, there are still uncertainties in the  distribution of column densities and Doppler {\it b} parameters of intervening absorbers, fundamental ingredients in the simulations.
One connected aspect which could also impact the dispersion in IGM transmission is that one expects that the sources responsible for IGM absorption are clustered in space. Gas clouds producing the absorption are connected to large scale structures and follow the general clustering of matter (Prochaska et al. 2014). This aspect remains to be fully treated in IGM simulations at z$>$2.5 and could affect both the mean transmission and its dispersion (Bershady et al. 1999, Inoue et al. 2014, Prochaska et al. 2014). These effects could likely lead to significant cosmic variance around the mean IGM extinction in models and modelling them would need to be fully tested from a number of independent LOS in large cosmological simulations. 

\begin{figure}[h!]
\centering
\includegraphics[width=9cm,height=9cm]{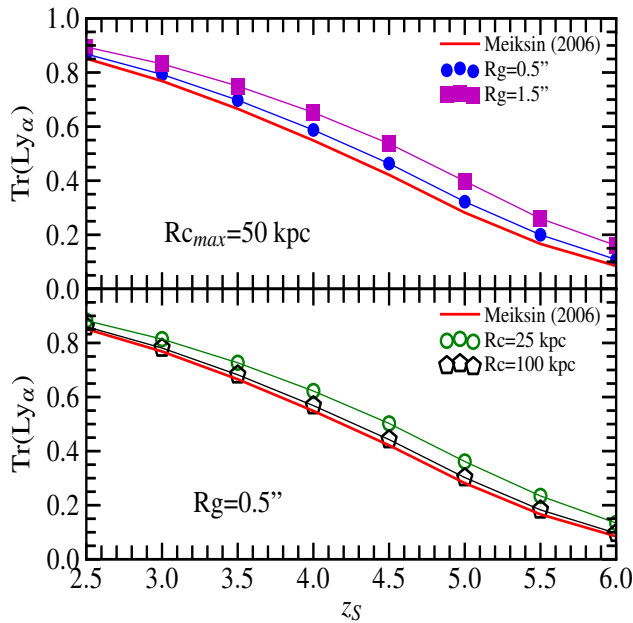}
\caption{Simulation of the IGM transmission expected when the source is an extended galaxy rather than a point-like QSO, ({\it Top panel}:) Simulation for different galaxy size from 0.5\arcsec to 1.5\arcsec and a maximum cloud size of 50 kpc. ({\it Bottom panel}). Simulation for a galaxy of 0.5\arcsec and different maximum cloud size of 25 and 100 kpc. The IGM transmission increases when the size of the clouds gets smaller, or when the size of the sources gets larger.}
\label{simul}
\end{figure}

An effect which can further complicate the computation of the mean IGM transmission and its dispersion in models is the covering factor of IGM clouds on an extended galaxy compared to a point source QSO. The smaller the clouds are, the more likely it is that the IGM clouds would block only part of an extended source. We simulate this effect taking a distribution of clouds, along the LOS of an extended galaxy with increasing size. This is shown in Figure \ref{simul}. The top panel represents the influence of the galaxy size (0.5\arcsec and 1.5\arcsec) on the IGM transmission for a maximum cloud size of 50 kpc. We find that when the size of the galaxy increases the transmission increases for a maximum cloud size fixed (top panel). This is in agreement with what we expect because only a part of the galaxy light is attenuated by the clouds along the line of sight. In this model the mean transmission is equal to that of QSOs for galaxy sizes becoming point-like as for QSOs. The bottom panel presents the same study performed with different maximum cloud size (25 and 100 kpc). We see that the bigger  the clouds are the smaller the difference between M06 and our simulation is. This is due to the fact that when the size of the clouds increases the light of the galaxy is more likely to be absorbed and the difference between QSOs and galaxies tends to disappear.

\section{Consequences for the selection of high redshift galaxies}
\label{selec}

The observed range of IGM transmission at a fixed redshift may have a significant impact on selecting galaxies at high redshifts z$>$2.5. The classical way to select galaxies at these redshifts is to use the Lyman Break Galaxy (LBG) selection technique. The LBG selection criteria rely on the a priori knowledge of the average properties of the IGM coupled to the 912\AA ~Lyman continuum limit (break) intrinsic to a galaxy (Steidel et al. 1996) as well as to the dust content and Lyman continuum escape fraction (Cooke et al. 2014). These components are responsible for the drop-out in flux and the resulting strong change in colours with redshift when the break in the continuum produced by the combination of the Lyman break and the IGM goes through a set of filters.  These properties are used to infer the locus of galaxies with different redshifts in colour-colour diagrams and to identify large samples of LBGs for  analysis of the galaxy population or for follow-up spectroscopic surveys (e.g. Steidel et al. 2003) at increasingly large redshifts.  The technique of deriving photometric redshifts from SED fitting also assumes average IGM properties (most often M95 in existing codes), and therefore is also expected to be dependent on the exact IGM transmission and its dispersion. As LBG and photometric redshift techniques have become the main way to identify large samples of candidate high-redshift galaxies to study galaxy evolution, it is of major importance to ensure that the assumed IGM properties are validated by observational data and that no major population is missed when using modelled average IGM properties a priori. 

\begin{figure}[h!]
\centering
\includegraphics[width=8cm,height=8cm]{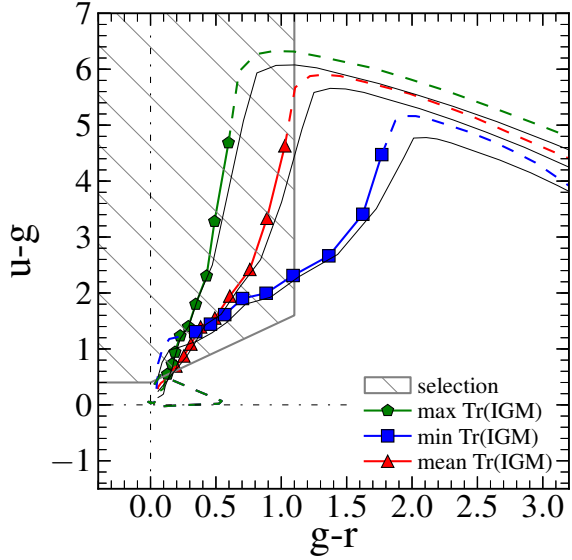}
\caption{Effect of the change in IGM transmission on the colour-colour (u-g,g-r) tracks of a galaxy with $2.5<z<3.5$ and its selection. As an example we use here a template galaxy with average properties at this redshift: $0.4$ Gyr old, with E(B-V)$=0.1$, a metallicity $Z_{\odot}/2.5$, and a SFH timescale of $0.3$ Gyr. The blue curve represents the (u-g,g-r) track for a  low IGM transmission (mean-1.0$\sigma$) while the green curve represents the track computed with a high IGM transmission  (mean+1.0$\sigma$). The red track is computed with the mean IGM transmission from M06. The grey dashed box corresponds to the LBG selection for this filter set (from Le F\`evre et al, 2013b). For each of the high or low transmission tracks the full line with squares (displayed at a step of $\Delta z=0.1$) follows from $z=2.7$ to $z=3.5$ , while the dashed line is for $z<2.7$ and $z>3.5$. We use an apparent (observed) flux below the Lyman limit of 3\% of the flux at 1500\AA ~as measured by Le F\`evre et al (2014) in the VUDS survey (see also Cooke et al. 2014). The black lines are the corresponding tracks computed with the M95 models.  
}
\label{SelectionUGR}
\end{figure}

We have modelled how galaxy tracks  in colour-colour space are modified when the IGM departs from the mean values of M06 and M95.  In Figure \ref{SelectionUGR} we present the tracks as a function of redshift of a galaxy in the (u-g,g-r) colour-colour diagram classically used as a basis for LBG galaxy selection at $z\sim3$. When the IGM transmission is low, the magnitude difference in bands straddling the continuum in the IGM-affected wavelength domain blueward of Ly$\alpha$ and redward of the Lyman limit is more pronounced than when the transmission is high. This has a strong impact on the (g-r) colour getting redder as the IGM transmission is decreasing, while the (u-g) colour changes less drastically as it is dominated by the Lyman-limit continuum break. The net resulting effect is that galaxies will leave the LBG box at lower redshifts when the IGM transmission is decreasing. For the simulated galaxy in  Fig. \ref{SelectionUGR} the tracks leave the selection box at $z\geq3.2$ instead of $z\geq3.5$ for our IGM model $-1\sigma$ below the mean.   
Interestingly this $mean-1\sigma$  track gets in the LBG selection box at $z=2.1$, earlier than $z=2.7$ as is expected from the mean transmission, which then implies that galaxies would be selected at lower redshifts than the anticipated $z=2.7$ limit. 
These effects are further compounded with the effect of a varying Lyman continuum escape fraction, as discussed in Cooke et al (2014). Significant flux is observed below the Lyman limit for high redshift galaxies, as discussed in Cooke et al. (2014) and observed in the VUDS sample discussed here with a flux $\sim3$\% of the flux at 1500\AA  (Le F\`evre et al. 2014). This {\it observed} flux below 912\AA ~could either be escaping Lyman Continuum photons from the galaxy or be artificially produced by contamination along the LOS, but either of these effects will affect the strength of the {\it observed} Lyman-limit break. In addition, photometric errors on the  (u-g) and (g-r) colours range from about 0.2 to 0.5 magnitudes and will further scatter objects outside of the selection box especially for objects near the boundaries of the box and therefore concerning objects for which IGM transmission along the line of sight is low.  

\begin{figure*}[h!]
\centering
\includegraphics[width=18cm,height=8cm]{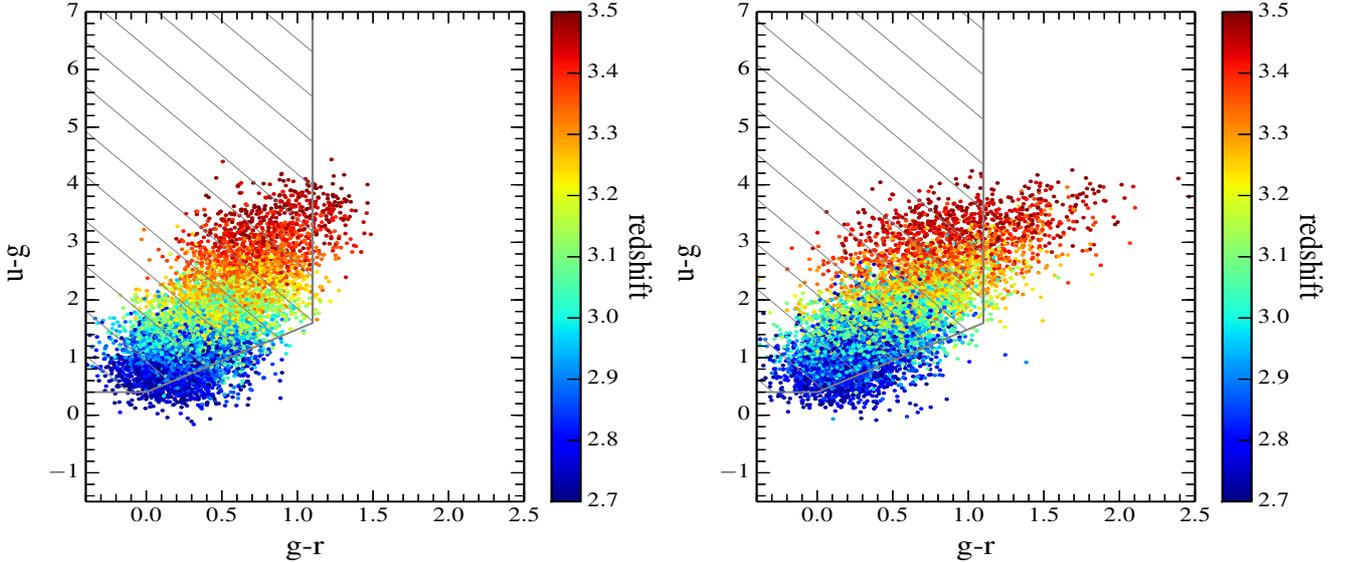}
\caption{The (u-g,g-r) colour-colour diagrams for galaxies simulated with BC03 synthetic galaxy spectra using BC03. ({\it Left}): simulation using a fixed IGM value at a given redshift following the mean value of the M06 model. ({\it Right}) simulation using a varying IGM transmission following the observed dispersion from QSOs of the VUDS sample presented here. }
\label{ugr_simul}
\end{figure*}

To estimate the efficiency of colour-colour selection in the presence of variable IGM transmission, we produce two large simulations of galaxies with 2.7$<$z$<$3.5 using the BC03 stellar population models, and a range of E(B$-$V), SFH, ages, and metallicity as defined in Table \ref{Paramspace}. The first simulation is based on a fixed M06 IGM transmission, while the second simulation randomly adds a variable IGM transmission with the 7 templates defined in Section \ref{igmpresc} following the observed dispersion. We derive the (u-g) and (g-r) colours for each simulated galaxy for the two simulations. The resulting (u-g,g-r) colour-colour plots are shown in Figure \ref{ugr_simul}, where is can be seen that when the IGM is  in the range reported in Section \ref{results}, a number of galaxies are out of the LBG selection box. We compute the relative fraction of galaxies coming in and out of the LBG selection box for this redshift range. We find that the fraction of objects outside of the LBG box increases to 16\% when the IGM is let to vary. This effect is more pronounced in the high redshift part of the selection. Between $z=3.2$ and $z=3.5$ the proportion of selected/unselected galaxies goes from 6\% to 23\%, just by making the IGM a free parameter.

\begin{figure}[h!]
\centering
\includegraphics[width=8cm,height=8cm]{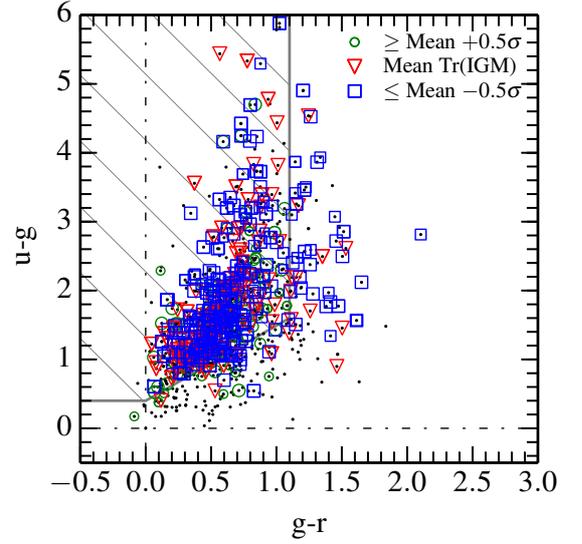}
\caption{The (u-g,g-r) colour-colour diagram for VUDS galaxies used in this study. The points are colour-coded based on the IGM transmission measured from the spectral fitting on each of them. The red points are the galaxies for which GOSSIP+ identified an IGM equal to the mean M06 model. The blue points are the objects for which the IGM transmission is -0.5, -1.0 or -1.5$\sigma$ below the mean, and green points are for +0.5, +1.0 or +1.5$\sigma$ above the mean. The galaxies which are outside the box have an IGM transmission below the mean M06 IGM transmission. Black points represent all the VUDS sample at 2.7$<$z$<$3.5.}
\label{ugr_igm}
\end{figure}

To check whether the observed sample agrees with the simulations, we plot the individual galaxies in the (u-g,g-r) diagram coded as a function of the IGM transmission measured from the spectral fitting in Figure \ref{ugr_igm}. The galaxies outside of the selection area for redshifts $2.7<z<3.5$ are predominantly galaxies for which the IGM transmission is lower than the mean, in agreement with simulations. Finding galaxies in the right redshift range but outside of the LBG selection area has been reported before (e.g. Le F\`evre et al. 2005b, Le F\`evre et al. 2013b); this paper shows that the large range in IGM transmission is one possible explanation for these observations.

\begin{figure}[h!]
\centering
\includegraphics[width=8cm,height=8cm]{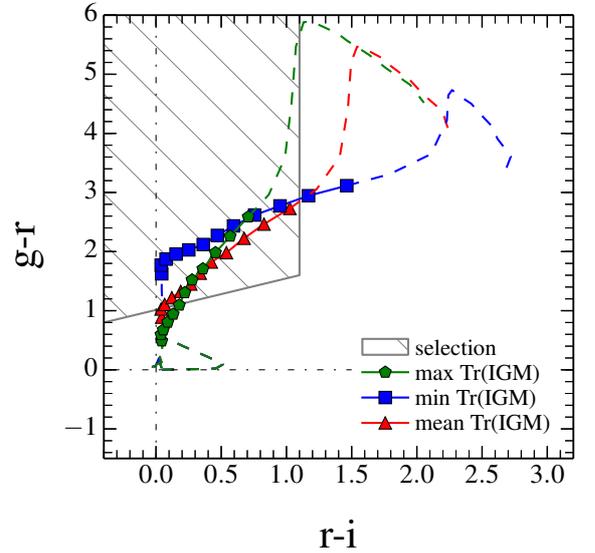}
\caption{Same as figure \ref{SelectionUGR} but for (g-r,r-i) colour-colour selection at 3.7$<z<$4.5.}
\label{SelectionGRI}
\end{figure}

Figure \ref{SelectionGRI} shows the same analysis for (g-r,r-i) colour-colour selection used to select galaxies between $z=3.5$ and $z=4.5$. The discussion for ugr selection also applies at these redshifts. The minimum IGM transmission allows to select galaxies as early as $z=3.1$ but limits the selection of the highest redshift galaxies to $z\leq4.3$ when the IGM transmission is lower than the mean transmission. 

From this analysis, we conclude that it is advisable to allow for a range of IGM transmission when defining the locus of galaxies in a colour-colour diagram for LBG selection, or when performing SED fitting to derive a photometric redshift (and associated physical parameters), prior to define large complete galaxy samples. This will be studied in more details in a forthcoming paper (Thomas et al. in prep.).

\section{Summary}
\label{summary}

This paper presents a study of the transmission of the inter-galactic medium using galaxy spectra over the redshift range $2.5<z<5.5$. We compute the IGM transmission from the highest SNR UV rest-frame spectra of 2127 galaxies with $i_{AB} \leq 25$ in the Vimos Ultra Deep Survey (VUDS, Le F\`evre et al. 2014) which uses integration times of $\sim14$h on VLT-VIMOS. 

To study the IGM transmission on the line of sight of these galaxies we use the GOSSIP+ software to perform spectral fitting of the VUDS VIMOS spectra. A new IGM prescription, which allows to let the IGM be a free parameter, is presented and implemented in the spectral fitting process. The IGM transmission can be any of seven IGM templates at any redshift defined from the mean M06 transmission and ranging up to $\pm1.5 \sigma$ as defined in the M95 models. We simulate mock galaxies to test the reliability of GOSSIP+ to retrieve the right IGM transmission for spectra with the same SNR as our observed sample and we measure the success rate to be $\sim74$\%. 
We then measure the IGM transmission by fitting our VUDS spectra, estimating this transmission directly on the  IGM template selected from the best fit.

The mean IGM transmission is found to be in remarkable agreement with the M06 theoretical model and measurements using QSOs up to z$=$4. We transform our transmissions into HI effective optical depths $\tau_{eff}$ and find that the mean values are also in excellent agreement with the  optical depths measured from the LOS towards QSOs as reported in the literature (e.g. Becker et al. 2013).  
The fact that the IGM transmissions derived from galaxies and from QSOs are in such excellent agreement in this redshift range seems to indicate that the source properties and the properties of the CGM in the immediate surroundings of these objects have little effect of the integrated IGM transmission properties, as would be generally expected.

We report tentative evidence that at z$>$4 the mean IGM transmission is higher than expected from the IGM model of M06, reaching a difference of $\sim$9\% at z=4.75. We discuss possible observational biases that could produce this difference and we conclude that these would not affect the sample in a way to produce a transmission observed to be higher than the mean M06 or than the values from QSOs, but rather would produce the opposite. We find that, not surprisingly,  at z$>$4 the dust extinction at the source is somewhat degenerate with the IGM transmission. The IGM is able to compensate a change in the E(B$-$V) value when performing the spectral fitting: the higher the dust content, the higher the IGM transmission has to be. When the E(B$-$V) is free to take values up to E(B$-$V)=0.5 we find a higher transmission than observed from QSOs, but when we restrict the dust extinction to E(B$-$V)$\leq$0.05 we find a mean IGM transmission compatible with the observed values from QSOs and simulations within measurement errors.  

Most importantly, we find that the dispersion of IGM transmission around the mean is large, ranging from  $1\sigma$=0.15 
at redshift $2.8$ to $1\sigma$=0.10 at redshift $4.8$. 
Our transmission values show a range and dispersion very similar to that reported by Faucher-Gigu\`eres et al. (2008b) from QSOs, confirming that our measurements are a true indication that the IGM dispersion is large at any of the redshifts observed.  
 We discuss the possibility that the high dispersion found in our sample may be the result of incomplete treatment of the clustering of IGM clouds in models resulting in large cosmic variance. We also discuss the possibility that the IGM transmission may be different when observing extended galaxies rather than point-like QSOs and present a simple model supporting this view.

The large dispersion in IGM transmission observed in our study has important consequences.
We explore the impact of this large dispersion on the selection of high redshift galaxies $z>2.5$ when using photometric techniques like LBG or photometric redshifts. We show that, when the IGM is allowed to vary, this changes the range of redshifts satisfying colour-colour (LBG) selection, and that part of the galaxy population in the range expected from the LBG selection technique can easily escape photometric selection. This is further compounded when IGM properties are combined to {\it observed} non zero flux below the Lyman limit, either from Lyman continuum escape fraction or from contamination along the LOS (see e.g. Cooke et al. 2014). We use large simulations to compare the colour-colour distribution when the IGM has a fixed value and when the IGM transmission is varying. We find that the fraction of galaxies which can escape detection in the LBG box may reach up to 16\% (23\% at 3.2$<z<$3.5) due to the IGM dispersion only. This fraction can further increase when a large range of IGM transmission is combined with a significant continuum level below 912\AA ~produced either by high Lyman continuum escape or by contaminating objects on the LOS. This will be explored with more extensive simulations (Thomas et al. in prep.). 

We have presented the first comprehensive study of the IGM transmission properties using a galaxy sample up to $z\sim5$. It provides a check of the IGM transmission properties independent from QSO samples. 
Our results demonstrates that the spectra fitting method we have developed including varying IGM transmission is fully able to recover the mean IGM transmission in the universe from a faint galaxy sample. Applying this powerful method to even larger galaxy samples will complement the knowledge of the IGM recovered from QSO spectra analysis and opens a new window of investigation for the IGM. 
While the average IGM transmission found in this work is in excellent agreement with QSO studies and models at z$<$4, the large distribution of  transmission observed in our sample and a possible higher IGM transmission at z$>$4 calls for more investigations, both from the observational and model perspectives, to better understand the IGM transmission properties.   

\begin{acknowledgements}
We thank G. Becker, J.M. Deharveng, A. Inoue, C. P\'eroux, and M. Pieri for useful discussions.  
This work is supported by funding from the European Research Council Advanced Grant ERC-2010-AdG-268107-EARLY and by INAF Grants PRIN 2010, PRIN 2012 and PICS 2013.
AC, OC, MT and VS acknowledge the grant MIUR PRIN 2010--2011.
This work is based on data products made available at the CESAM data center, Laboratoire d'Astrophysique de Marseille, France.
\end{acknowledgements}

\end{document}